\documentclass[11pt]{article}
\usepackage{fullpage,amssymb,amsmath,graphicx,natbib,multirow,hyperref}

\renewcommand{\v}[1]{\boldsymbol #1}

\title{Comment on ``Bayesian Nonparametric Inference - Why and How'' by M\"uller and Mitra}
\author{ 
Peter Hoff \\ 
Departments of Statistics and Biostatistics\\
University of Washington }
\date{\today}

\begin{document}
\maketitle

\begin{abstract}
Due to their great flexibility, 
nonparametric Bayes methods have proven to be a valuable tool 
for discovering complicated patterns in data. 
The term ``nonparametric Bayes'' suggests that these methods 
inherit model-free operating characteristics of 
classical nonparametric methods, as well as coherent 
uncertainty  assessments provided by Bayesian 
procedures. 
However, 
as the authors say in the conclusion to their article, 
nonparametric Bayesian methods may be more aptly described as 
``massively parametric.''  
Furthermore, I argue that many of
the default nonparametric Bayes procedures 
are  only Bayesian in the weakest sense of the term, 
and
cannot be assumed to provide honest assessments of 
uncertainty merely because they carry the Bayesian label.  
However useful such procedures may be, 
we should be cautious about advertising default nonparametric Bayes 
procedures as  either  being ``assumption free''  or 
providing descriptions of our uncertainty. If we want our 
nonparametric Bayes procedures to have a Bayesian interpretation, 
we should modify default NP Bayes methods to accommodate 
real prior information, or at the very least, carefully evaluate 
the effects of hyperparameters on posterior quantities of interest.

\end{abstract}

\section{Parameteric and nonparametric approaches}
Historically, a standard justification of Bayesian methods 
has been that they provide an internally consistent approach to 
updating information:
If  $\mathcal P_{\Theta} = \{ p(y|\theta):\theta\in \Theta\}$
expresses
our beliefs about $Y$ given $\theta$, and
$\pi(\theta)$ expresses our beliefs about $\theta$,
then $\pi(\theta|y) \propto \pi(\theta) p(y|\theta)$ expresses
what we \emph{should} believe about $\theta$, having observed $Y=y$.
From this subjective Bayesian point of view, for $\pi(\theta|y)$ to be of
most use, both
$p(y|\theta)$ and $\pi(\theta)$ should actually represent our
beliefs, at least approximately.
A criticism of parameteric Bayesian methods is 
that commonly used  models $\mathcal P_\Theta$  are often suspected 
of being wrong. 
Nonparametric Bayes methods appear to 
solve this problem by making $\mathcal P_\Theta$ so large that it 
 includes  essentially all relevant sampling distributions $p(y|\theta)$.  
The authors seem to suggest that NP Bayes methods therefore 
provide an
``honest representation of uncertainties''.
I would agree with this, to the extent that 
$\pi(\theta)$ actually represents prior beliefs.

How honest are parametric and nonparametric priors?
Parametric priors are arguably inaccurate
as they assign probability one to a simple parametric model. 
However, the great advantage of parametric Bayesian approaches
is that they allow the prior to be specified in terms of
parameters of interest, which often happen to be the parameters about which
we have real prior information.
As a very simple example,
suppose we have a sample $y_1,\ldots, y_n$ of
independent observations from a population for which
we have prior information about the mean and variance.
In this case, a limited form
of subjective, robust Bayesian inference for the population mean $\theta$
can proceed via
the posterior density obtained from a normal sampling model
for $y_1,\ldots, y_n$.
While the likelihood may not be exactly correct,
the resulting inferences for $\theta$
are robust to nonnormality, asymptotically correct and
provide confidence intervals for which the
asymptotic frequentist coverage equals the asymptotic
Bayesian coverage.
Perhaps most importantly, the inference is transparent:
the approximate normality of $\bar y$ is well understood,
and the effect of the prior on the parameter is simple,
especially if a conjugate prior is used. Even if the prior
does not represent our actual prior information, at least we
can understand what information it represents.

In contrast, I think it is safe to say 
that for most NP Bayes methods used in practice, 
the prior does not represent actual prior beliefs, 
even 
approximately. 
One difficulty is that standard NP Bayes priors 
include hyperparameters that 
directly control things that we are unlikely to 
have prior information about (the number of modes of a density)
and only indirectly control things we might have information about 
(means, variances and correlations). 
For example, the choice of 
the hyperparameters in the prior for a  Dirichlet process mixture model (DPMM) 
induces 
a prior on the mean and variance of the population, 
but the mapping 
from the hyperparameters to these induced priors can be very opaque 
\citep{yamato_1984,lijoi_regazzini_2004}.  
Similarly, the P\'olya tree priors discussed in Section 2.2 
require the specification of a partition over the sample  space, 
the choice of which will generally affect the posterior. 
The ``solution'' to this is the addition of a  prior 
over the set of possible partitions. 
It is hard to imagine that such a prior represents actual 
prior information about the underlying population. 

Do such complications warrant abandoning NP Bayes methods and 
using simpler parameteric approaches?  It may depend on the 
data analysis objectives.  
NP Bayes methods provide a flexible means of representing 
high-dimensional data structure. Overfitting is avoided by 
a regularizer (the prior) that has a probabilistic 
interpretation. These features 
make NP Bayes methods an attractive set of tools 
for such tasks 
as  prediction and clustering. 
However, if the data analysis objective is to 
describe our 
posterior 
information about a parameter of interest, then the 
appropriateness of NP Bayes is less clear.  
Example 1 from 
\citet{muller_mitra_2013}
is a situation 
where the use of an NP Bayes method
may be obfuscating the sources of information about the 
parameter interest, $F(0)$. 
If we are to take the likelihood 
at face value, 
then it seems the data have little to say about the value of 
$F(0)$:
Writing $f_k=F(k)$ and letting $n_k$ be the number 
of $T$-cell sequences for which $k$ replicates 
were observed,  
the likelihood can be expressed as $p(\v y | f_0,\ldots, f_4 ) =\prod_{k=1}^4 ( f_k/[1-f_0])^{n_k}$. 
How much information do the data provide about $f_0$? 
One way to evaluate this is to consider the 
profile likelihood function of $f_0$. 
For every fixed value of $f_0$ the 
likelihood 
$p(\v y | f_0,\ldots, f_4 ) =\prod_{k=1}^4 ( f_k/[1-f_0])^{n_k}$
is maximized in $f_1,\ldots, f_4$ at 
$\hat f_k = (1-f_0) n_k/n $. 
This gives a constant profile likelihood function for $f_0$, 
equal to 
 $\prod ( n_k/n)^{n_k}$ for every value of $f_0$.
From a Bayesian perspective, 
the posterior 
distribution of 
$f_0$  can be expressed as   
$\pi( f_0 | \v y)   \propto  \pi(f_0) p(\v y|f_0) $, where 
\begin{align*}
 p(\v y|f_0) & = \int p(\v y | f_0,\ldots, f_4 ) \pi(f_1,\ldots, f_4|f_0) 
   \ df_1 \cdots df_4 . 
\end{align*}
The profile likelihood argument suggests that $p(\v y|f_0)$ 
will be fairly flat as a function of $f_0$, especially if the prior
over $f_1,\ldots, f_4$ is ``diffuse,'' 
in which case $\pi( f_0 | \v y)  \approx \pi(f_0)$. 
In this case, an ``honest assessment'' of posterior uncertainty about 
$f_0$ requires only an honest prior for $f_0$.  
Alternatively, if we believe there to be a relationship 
among $f_0,\ldots, f_4$ beyond the fact that $f_0+\cdots +f_4<1$, then again an honest assessment of 
posterior  uncertainty about $f_0$ for these data requires only 
an honest specification of ones joint beliefs about the 
five numbers $f_0,\ldots, f_4$.
In either case, 
the  DPMM
over $\{ f_0,f_1,\ldots \}$ seems like a very indirect and 
opaque way to specify a prior over the relevant parameters. 

One could argue that the DPMM in this example is helping 
us estimate other aspects of the unknown frequencies, such as 
the relative frequencies, perhaps.  However, even though the DPMM 
in this example is billed as a ``nonparametric Bayes'' procedure, it is not 
without strong modeling assumptions. Specifically,
this DPMM assumes the true distribution is a mixture of  
Poisson distributions  - a class of distributions that does not
contain all discrete distributions.

\section{Partial remedies for particular situations}

\paragraph{Alternative likelihoods:}
Consider a model $\{ p(y|f) : f\in \mathcal F\}$ where 
$f$ is a high dimensional parameter (such as a regression function 
or density).
In situations where primary interest is in a  
low-dimensional parameter $\theta=\theta(f)$, 
the difficulties of infinite-dimensional prior specification
can sometimes be avoided by using a likelihood that involves
only $\theta$. 
For example, in 
many problems there exists a  
 statistic $t(y)$ whose distribution depends only on $\theta$,
and not the high-dimensional parameter $f$. In such 
cases, the 
likelihood can be expressed as 
\[ p(y | f ) = p( t(y) | \theta ) \times p( y| t(y),\theta, f). \]
The need to specify a prior over $f$ can be avoided 
by constructing a posterior distribution for $\theta$ 
based only on the marginal likelihood $p(t(y)|\theta)$, i.e.\ 
$\pi( \theta | t(y)) \propto \pi(\theta) p(t(y)|\theta)$. 
Estimates based on such a posterior distribution could be
inefficient, as they ignore any additional information about 
$\theta$ in $p(y|t(y),\theta,f)$, but they do not 
require specification of a prior for the high-dimensional 
nuisance parameter $f$.
A concrete example of such a procedure
is given in \citet{hoff_2007a} in  the context of a 
semiparametric copula model, in which
$\theta$ represents the parameters in a parametric
dependence model and $f$ parameterizes a set of unknown
infinite-dimensional univariate marginal distributions.

Many researchers have considered other alternative likelihoods 
for robust or ``nonparametric'' Bayesian inference 
(\citet{efron_1993, lazar_2003, greco_racugno_ventura_2008}, to name a few).
Asymptotically correct likelihoods 
for parameters of interest
can even be derived from 
misspecified models: Very generally, the 
limiting distribution of the MLE $\hat \theta$ in a misspecified 
model is asymptotically normal, so that 
\[ \sqrt{n} (\theta^* - \hat \theta )  \stackrel{\cdot}{\sim} N( 0 , V(\theta^*) )  \]
where $\theta^*$ is the
``pseudotrue'' parameter and $V(\theta^*)$ is the ``sandwich'' variance
\citep{huber_1967}. 
In many cases (such as in exponential family models) 
 $\theta^*$  is a population moment, and possibly the parameter 
of interest about which we may have prior information. In this case, 
nonparametric Bayesian inference can be obtained by 
combining 
a prior on $\theta^*$ with 
the asymptotic normal distribution of $\hat \theta$ as a likelihood. 
Such ``Bayesian sandwich'' procedures have been considered 
\citet{szpiro_rice_lumley_2010}, \citet{muller_2012} and 
\citet{hoff_wakefield_2012_tr}.

\paragraph{Marginally specified priors:} 
Such reductions of the parameter space are not feasible in
applications such as prediction, where
the high- or infinite-dimensional  parameter $f$ is of primary interest.
In such cases it is important from a Bayesian perspective that
the prior for $f$ reflects known information as much as possible. 
Realistically, a statistician is unlikely to have informed opinions about all aspects of a  high-dimensional parameter $f$, 
but may have real information about 
a finite-dimensional functional $\theta=\theta(f)$, 
such 
as the population mean or variance. 
Recently, 
\citet{kessler_hoff_dunson_2012_tr} 
have proposed an approach for incorporating prior information 
about $\theta$ into a default  NP Bayes prior for 
$f$. Specifically, let  $\pi_0$ be a prior for $f$ that is 
chosen arbitrarily or for computational convenience.
This prior induces a marginal 
prior on $\theta$, say $P_0$,  that may not reflect 
actual prior information, as quantified by a distribution  $P_1$.  
To remedy this problem, first 
express the default prior $\pi_0$ as 
\[  \pi_0( f\in A) =\int  \pi_0( f\in A |\theta) P_0(d\theta). \]
To obtain a prior $\pi_1$ over $f$ with the desired marginal distribution 
$P_1$, simply replace $P_0$ with $P_1$ in the above expression. 
 The resulting prior on $f$ then becomes
\[  \pi_1( f\in A) =\int  \pi_0( f\in A |\theta) P_1(d\theta). \]
Such a prior generally retains the good large-support properties of 
the default NP Bayes prior $\pi_0$, but has an induced prior 
over $\theta$ that matches the actual prior information $P_1$.  
Computation of the posterior under such a prior is also 
generally available if an MCMC algorithm exists for the 
default prior $\pi_0$.  In this case, posterior approximation 
under $\pi_1$ can be made with the addition of a Metropolis-Hastings 
step. 

\paragraph{Noninformative priors:}
In the absence of prior information, NP Bayes practitioners 
may
attempt to produce ``diffuse'' priors by adjusting the  
hyperparameters in some way.  
However, intuition about what parameters correspond to 
``noninformativeness'' can be misleading, partly due to the 
terminology used in the NP Bayes literature.
For example, 
NP Bayes researchers should make clear that  the
total mass parameter $\alpha$  
in a DPMM
controls much more than ``the uncertainty
of'' the mixing measure.  As DPMM researchers know, 
this hyperparameter controls such things as the entropy of the resulting
probability density, the number of modes, etc.
As another example, practitioners sometimes 
select overdispersed base measures in DPMMs
in the hope that these reduce the effect of the prior on the analysis.
\citet{bush_lee_maceachern_2010} have shown that such attempts generally
lead to an unreasonably small number of mixture components,
and propose a more nuanced version of the total mass hyperparameter to
achieve a type of ``noninformative'' NP Bayes analysis.

\section{Conclusion}
Standard data analysis procedures can generally be described
as techniques that convert a large set of numbers (the data)
into a smaller set of numbers (parameter estimates, standard errors, 
etc.). Ideally, such a procedure is
statistically meaningful and reasonably transparent:
meaningful in that it  has some desirable
property in an idealized situation,
 and transparent so that its  behavior can be understood 
outside of the idealized situation. 
Conjugate Bayesian estimation 
in an exponential family model is a good example 
of a meaningful and transparent procedure: The properties of such procedures 
are well-understood in the subjective Bayes framework, in an asymptotic 
framework and even in settings where the model is misspecified. 
In many cases, 
incorrect parameteric models can provide
meaningful, transparent
and accurate inference for certain  population parameters of interest, if not
for all aspects of the population.

NP Bayes procedures 
typically convert a small set of numbers (the data) into 
a much larger set of numbers (the posterior distribution of the 
infinite dimensional parameter). 
What meaning can such  extrapolative procedures have? 
Asymptotic  results 
assure us 
that many NP Bayes procedures converge to the truth
as fast as other nonparametric procedures, and perhaps faster
if the the prior is close to the truth in a topological sense
(see, for example, \citet{ghosal_2001}).  It might even be the
case that 95\% posterior confidence intervals have  approximate 95\%
frequentist coverage, as they often do in parametric models
\citep{severini_1991}.

How are NP Bayes procedures justified non-asymptotically?
Small sample justifications of Bayesian procedures 
are often based on their optimality under a particular prior.
In simple models, this justification is transparent, in that 
even if the prior doesn't represent one's actual beliefs, 
at least one understands what beliefs it represents.  
Prior specification for 
NP Bayes procedures
are more opaque and 
harder to justify from a Bayesian perspective. 
Default prior distributions are not generally going to represent prior beliefs,
making it difficult to interpret the corresponding 
posterior distributions as posterior 
beliefs (except perhaps asymptotically). 
In terms of transparency, it is certainly possible to gain 
a strong intuition for the effects of hyperparameters on 
posterior output. Yet expert nonparametric Bayesians frequently use 
terminology that may be misleading to 
less experienced NP Bayes practitioners: 
For example, referring to the mass parameter $\alpha$ in a DPMM as  
indexing uncertainty is an incomplete description at best. 
Referring to a DPMM as a method for ``BNP inference'' on a clustering 
overlooks, as the authors pointed out in Section 3.1, that 
the P\'olya urn scheme is a very particular one-parameter partition model. 
Referring to a mixture of Poisson distributions as ``nonparametric''
may give the impression to the inexperienced reader that the resulting mixture 
model contains all discrete distributions. 

Most importantly, a posterior distribution does not provide an 
honest assessment of uncertainty by virtue of being a posterior 
distribution.  Such an assessment is obtained via either 
an honest prior  or asymptotically. In the absence of an infinite 
sample size, considerable effort should be made to use a prior distribution 
that 
approximates as closely as possible any real prior information that 
is available. 
In the absence of prior information, a more complete (but tedious)
description of uncertainty would include a sensitivity analysis 
over possible values of the hyperparameters. 

I am certainly not arguing that such efforts regarding the 
prior be mandated for every application of 
an NP Bayes method. However, I feel that more effort in this direction 
is necessary if we want our posterior distributions to represent 
honest assessments  of uncertainty.

\bibliographystyle{chicago}
\bibliography{/Users/hoff/Dropbox/SharedFiles/refs}

\end{document}